\documentclass[]{pasj01}
\begin{document} 
\Received{}
\Accepted{}

\title{An Analytical Density Profile of Dense Circumstellar Medium in Type II Supernovae}

\author{Daichi \textsc{Tsuna}\altaffilmark{1,2}
\email{tsuna@resceu.s.u-tokyo.ac.jp}}
\altaffiltext{1}{Research Center for the Early Universe (RESCEU), School of Science, The University of Tokyo, 7-3-1 Hongo, Bunkyo-ku, Tokyo 113-0033, Japan}
\altaffiltext{2}{Department of Physics, School of Science, The University of Tokyo, Tokyo, Japan}
\author{Yuki \textsc{Takei}\altaffilmark{1,3,4}
}
\author{Naoto \textsc{Kuriyama}\altaffilmark{1,3}
}
\author{Toshikazu \textsc{Shigeyama}\altaffilmark{1,3}
}
\altaffiltext{3}{Department of Astronomy, School of Science, The University of Tokyo, Tokyo, Japan}
\altaffiltext{4}{Astrophysical Big Bang Laboratory, RIKEN, 2-1 Hirosawa, Wako, Saitama 351-0198, Japan}

\KeyWords{circumstellar matter --- supernovae: general --- stars: mass-loss --- radiative transfer} 

\maketitle

\begin{abstract}
Observations of Type II supernovae imply that a large fraction of its progenitors experience enhanced mass loss years to decades before core collapse, creating a dense circumstellar medium (CSM). Assuming that the CSM is produced by a single mass eruption event, we analytically model the density profile of the resulting CSM. We find that a double power-law profile, where the inner (outer) power-law index has a characteristic value of -1.5 (-10 to -12), gives a good fit to the CSM profile obtained using radiation hydrodynamical simulations. With our profile the CSM is well described by just two parameters, the transition radius $r_*$ and density at $r=r_*$ (alternatively $r_*$ and the total CSM mass). We encourage future studies to include this profile, if possible, when modelling emission from interaction-powered transients.
\end{abstract}

\section{Introduction} 
Recent observations of hydrogen-rich supernovae (SNe) imply that at the end of a massive star's life mass loss of the star is likely to be greatly enhanced \citep{Yaron17,Morozova18,Bruch20}. An enhanced mass loss forms a massive circumstellar medium (CSM) around the star. This can leave a significant observational consequence to the SN that follows, as the presence of a dense CSM can efficiently convert the SN ejecta's kinetic energy to radiation. The efficiency is roughly $\sim M_{\rm CSM}/(M_{\rm ej}+M_{\rm CSM})$, where $M_{\rm ej}$ and $M_{\rm CSM}$ are the mass of the ejecta and the CSM, respectively (e.g., \cite{Murase19}). For a typical $M_{\rm ej
}=10\,M_\odot$, a CSM of $\gtrsim 0.1M_\odot$ can dissipate $\gtrsim 1$\% of the kinetic energy, and the resulting emission can outshine otherwise normal SNe with a typical efficiency of $1\%$. This CSM interaction is widely believed to be the origin of the optically bright transients such as Type IIn and some superluminous SNe (e.g. \cite{Grasberg87,Chevalier11,Moriya18}). Observations of these SNe are an important probe of the dramatic final stages of massive star evolution.

The density profile $\rho(r)$ of the CSM as a function of the radius $r$ is a key to interpret observational data of light curves in various wavelength. Nevertheless, most works focus on a power-law profile, often a density profile of $\rho(r)\propto r^{-2}$ analogous to steady winds (e.g. \cite{Moriya11,Chevalier11,Chatzopoulos12,Svirski12,Ginzburg12,Moriya_et_al_13,Tsuna19,Takei20,Suzuki20}), where $r$ denotes the distance from the center of the star. While this approximation is simple and may enable analytical modelling of the light curve (e.g. \cite{Chatzopoulos12,Moriya_et_al_13,Tsuna19}), its validity may be questionable.

Recently \citet{Kuriyama20a} studied the CSM created from eruptive mass loss, by injecting energy at the base of an evolved massive star's envelope and numerically following the envelope's response. They found that the mass of the CSM inferred from Type IIn SNe can be naturally explained if the injected energy is comparable to the binding energy of the envelope. However they find a CSM density profile close to a double power-law, with the inner part following roughly $\rho\propto r^{-1.5}$ (see also \cite{Tsuna20}), shallower than a steady wind profile. While this discrepancy is interesting to investigate, interpretation of this profile was not given in their work.  In this work we aim to explain this profile, and present a simple analytical formula of the density profile that can be adopted for future modelling efforts of interaction-powered SNe.

This paper is constructed as follows. In Section \ref{sec:Model} we derive our analytical modelling of the density profile, which includes various parameters that govern its shape and normalization. Then in Section \ref{sec:Results} we demonstrate that the analytical model sufficiently describes the profile obtained by a more rigorous numerical approach. In Section \ref{sec:Discussion} we discuss the consequences of our profile on the light curve of interaction-powered transients, and also present a guide for using our analytical profile. We conclude in Section \ref{sec:Conclusion}.

\section{Analytical Model}
\label{sec:Model}
We consider the density profile of a CSM made from a single mass eruption. There are models explaining CSM by continuous mass loss (e.g., \cite{Soker21}), which would result in different profiles from what is derived in this work.

If only a fraction of the envelope's binding energy is injected, the resulting CSM should have both bound and unbound components. 
The CSM will be pulled by the central star's gravity and the bound component will eventually fall back. The latter has not been inspected in \citet{Kuriyama20a}, and we consider it in detail here.

Since the pressure in the CSM is reduced to a negligible value in several dynamical timescales from mass eruption, only gravity from the central star controls the motion of the CSM. Then assuming that the CSM is spherically symmetric, we obtain the trajectory of a CSM element at radius $r$ as \citep{Kuriyama20a}
\begin{equation}
\left(\frac{dr}{dt}\right)^2=\frac{2GM_r}{r}+2E_0,
\label{eq:drdt}
\end{equation}
where $G$ is Newton's constant, $M_r$ is the enclosed mass at $r$, and $E_0$ is the total (gravitational + kinetic) energy of a given fluid element, which is negative for a bound CSM. In this work we focus on the density profile at the two limiting regimes, the bound (innermost) limit and the unbound (outermost) limit.

\subsection{Bound CSM}
To analytically consider the inner bound CSM we make the following three assumptions.\\
(i) The density profile of the CSM is continuous and smooth, so that its local derivatives can be well defined.\\
(ii) The radius where the CSM is launched ($r_0$), which is roughly the progenitor's radius, is much smaller than the radius we focus on. This is usually satisfied, although it can be marginal for RSGs with radii exceeding $10^3R_\odot$.\\
(iii) The CSM mass is much smaller than the mass of the central star. This makes $M_r$ in equation (\ref{eq:drdt}) independent of $r$, and one can approximate it by the value of the central star's mass $M_*$. While this assumption may be challenged for some superluminous SNe with a massive CSM, it generally holds for the cases considered in our work.

We parameterize a solution of equation (\ref{eq:drdt}) as
\begin{equation}\label{r-theta}
    r = \frac{GM_*}{(-E_0)}\sin^2\left(\frac{\alpha}{2}\right)\ (\alpha_0<\alpha<2\pi-\alpha_0),
\end{equation}
where $\alpha_0$ is the value of $\alpha$ when $r=r_0$. We take $\alpha$ to be increasing with time. The velocity is then obtained as
\begin{equation}
    v=\pm\sqrt{GM_*(1+\cos\alpha)/r}, 
    \label{eq:v_of_theta}
\end{equation}
where the sign is positive (negative) when $\alpha$ is smaller (larger) than $\pi$, and $\alpha=\pi$ corresponds to the point where the material begins to fall back. Using equations (\ref{eq:drdt}) and (\ref{eq:v_of_theta}), the time $t$ can be expressed as a function of $\alpha$ as
\begin{equation}
    t = \frac{GM_*}{(-2E_0)^{(3/2)}}[(\alpha-\sin\alpha)-(\alpha_0-\sin\alpha_0)],
\end{equation}
 when $\alpha$ increases with time $t$.
Under the assumption (ii) $\alpha_0\ll 1$, we can approximate $\sin^2(\alpha_0/2)\approx (\alpha_0/2)^2$ and $\alpha_0-\sin\alpha_0\approx \alpha_0^3/6$. Then
\begin{equation}\label{t-theta}
    t+\frac{1}{3}\sqrt{\frac{2r_0^3}{GM_*}} = \frac{GM_*}{(-2E_0)^{(3/2)}}(\alpha-\sin\alpha).
\end{equation}
Setting $t_0\equiv \sqrt{2r_0^3/GM_*}/3$, we eliminate $E_0$ from equations (\ref{r-theta}) and (\ref{t-theta}), and obtain
\begin{equation}
    \frac{r}{(GM_*)^{1/3}(t+t_0)^{2/3}} = \frac{1-\cos\alpha}{(\alpha-\sin\alpha)^{2/3}}.
\end{equation}
For a given $X\equiv r/[(GM_*)^{1/3}(t+t_0)^{2/3}]$, we can obtain $\alpha$ and $v/\sqrt{GM_*/r}$. This parameter $X$ is associated with the ratio of the gravitational fallback timescale
\begin{equation}
t_{\rm fb}\approx \sqrt{\frac{r^3}{GM_*}} \sim 0.9\ {\rm yr}\left(\frac{r}{10^{14}\ {\rm cm}}\right)^{3/2}\left(\frac{M_*}{10M_\odot}\right)^{-1/2}
\end{equation}
and the age $t$, with the relation $X\approx (t_{\rm fb}/t)^{2/3}$. Thus material with $X$ smaller than about 1 has already started to fall back, whereas material with $X$ greater than this has not started to fall back yet\footnote{The exact border where the fall back starts is found to be at $X\approx 0.93$.}.

We find that we can approximate $v/\sqrt{GM_*/r}$ as a function of $X$ as 
\begin{eqnarray}
    \frac{v(r,t)}{\sqrt{GM_*/r}} = -0.159X^3+1.594X^{3/2}-1.300 \equiv f(X).
    \label{eq:v_of_r_t}
\end{eqnarray}
The error of this approximation is within 0.05 for $X>0.1$.
The range of $X$ is $X_{\rm in, CSM}< X < 6^{2/3}/2$, where the upper limit is the value of $(1-\cos\alpha)/(\alpha-\sin\alpha)^{2/3}$ in the limit of $\alpha\to 0$, and the lower limit $X_{\rm in, CSM}$ is the value for the element at the inner edge of the CSM $R_{\rm in, CSM}$ at a given snapshot,
\begin{eqnarray}
X_{\rm in, CSM} \sim 0.1\left(\frac{R_{\rm in, CSM}}{500R_\odot}\right)\left(\frac{M_*}{10M_\odot}\right)^{-1/3}\left(\frac{t+t_0}{5\ {\rm yr}}\right)^{-2/3}.
\end{eqnarray}
The innermost radius of the CSM $R_{\rm in, CSM}$ is in general different from the orignial photospheric radius of the progenitor $r_0$. This is because after mass eruption the progenitor's surface oscillates, and can settle down to the orignial position only after about a Kelvin-Helmholtz timescale, which is at least decades for evolved massive stars. For the progenitors that we have adopted in this work, the difference is nevertheless within a factor of two (see Table \ref{table:progenitors} and \ref{table:results}).

Plugging $v=f(X)\sqrt{GM_*/r}$ to the continuity equation
\begin{eqnarray}
\frac{\partial \rho}{\partial t}+\frac{1}{r^2}\frac{\partial}{\partial r}(r^2\rho v) =0,
\end{eqnarray}
we convert it to a differential equation with respect to $r$ and $X$:
\begin{eqnarray}
-\frac{2}{3}\frac{X^{3/2}}{f}\frac{\partial \ln\rho}{\partial \ln X} + \frac{\partial \ln\rho}{\partial \ln r} + \frac{\partial \ln\rho}{\partial \ln X} + \frac{3}{2} + \frac{\partial \ln f}{\partial \ln X} = 0.
\end{eqnarray}
Comparing this with the total derivative of the density
\begin{eqnarray}
d\ln\rho = \frac{\partial \ln\rho}{\partial \ln r} d\ln r + \frac{\partial \ln\rho}{\partial \ln X} d \ln X,
\end{eqnarray}
we obtain
\begin{eqnarray}
\frac{d\ln\rho}{-\frac{3}{2}-\frac{\partial \ln f}{\partial \ln X}} = \frac{d\ln X}{1-\frac{2X^{3/2}}{3f}} = d\ln r.
\end{eqnarray}
Substitution of $f(X)$ in equation (\ref{eq:v_of_r_t}) into these equations yields
\begin{eqnarray}
\frac{d\ln\rho}{dX} &=& \frac{-4.5X^3+30.047X^{3/2}-12.247}{X^4-5.828X^{5/2}+8.165X}, \\
\frac{d\ln r}{dX} &=& \frac{X^3-10.016X^{3/2}+8.165}{X^4-5.828X^{5/2}+8.165X}.
\end{eqnarray}
We solve $\rho$, $r$ as a function of $X$. After some algebra
\begin{eqnarray}
\rho &=& FX^{-3/2}\frac{(3.485-X^{3/2})^{6.334}}{(2.343-X^{3/2})^{8.334}}\equiv Fg(X) \\
r &=& BX\left(\frac{2.343-X^{3/2}}{3.485-X^{3/2}}\right)^{2.445}\equiv Bh(X).
\end{eqnarray}
We note that $X^{3/2}< (6^{2/3}/2)^{3/2}\approx 2.12<2.343$. Since the initial conditions give a relation between $F$ and $B$ as $F=F(B)$, the general solution for $\rho$ can be written as
\begin{eqnarray}
\rho(r,X) = F\left(\frac{r}{h(X)}\right)g(X).
\end{eqnarray}
Although the exact functional form of $F$ is uncertain, we can still obtain the profile at the limit of small $r$ and $t\gg t_0$. For $X\to X_{\rm in, CSM}\ll 1$, we find that $r/h(X)$ becomes independent of $r$ and $g(X)\propto r^{-3/2}$. Thus regardless of the exact form of the function $F$, we expect that the profile follows $\rho\propto r^{-3/2}$ when we are looking at the region where $X\sim X_{\rm in, CSM}$. This asymptotic behaviour is analogous to spherical Bondi accretion \citep{Bondi52}, that solves the spherical steady-state accretion flow onto a central object. In fact these two settings are essentially the same in the region where $X\ll 1$, i.e. where $t_{\rm fb}\ll t$ and the steady-state assumption is valid.

\begin{table*}
\centering
\begin{tabular}{ccccccc}
Model & $M_{\rm ZAMS}\ [M_\odot]$ & $Z\ [Z_\odot]$ & $R$ [$R_\odot$] & $T_{\rm eff}$ [K] & $M_{\rm He, core}$ & $M_{\rm H, env}$ \\ \hline
R15 & 15 & 1 & 670 & 4000 & 4.9 & 7.9\\
R20 & 20 & 1 & 840 & 4000 & 6.7 & 11.6\\
B20 & 20 & $10^{-2}$ & 130 & 9900 & 8.9 & 11.0
\end{tabular}
\caption{Properties of the progenitor stars adopted in this work to calculate the numerical CSM profile. The first two are red supergiants, and the last is a blue supergiant. The columns are: model name, ZAMS mass, metallicity, radius, effective temperature, mass within helium core, and mass of hydrogen envelope (total mass minus $M_{\rm He, core}$).}
\label{table:progenitors}
\end{table*}

\subsection{Connection to the unbound CSM}
At large $r$, the effect of gravity is usually negligible and the density profile is expected to be similar to that of the outer ejecta in normal SNe. The profile changes according to whether the outer envelope is convective or radiative, but is generally steep, with $\rho\propto r^{-12}$ for red supergiants (RSGs) and $\rho\propto r^{-10}$ for blue supergiants (BSGs) \citep{Matzner99}. We consider an interpolation of the inner and outer power-law limits, with a harmonic mean inspired by \citet{Matzner99}
\begin{eqnarray}
\rho_{\rm CSM}(r) = \rho_* \left[\frac{(r/r_*)^{1.5/y}+(r/r_*)^{n_{\rm max}/y}}{2}\right]^{(-y)},
\label{eq:rho_CSM}
\end{eqnarray}
where $r_*, \rho_*$ are the transition radius and density, $n_{\rm max}=12\ (=10)$ for RSGs (BSGs), and $y$ is a parameter that controls the curvature (smaller $y$ gives sharper transition). 

\section{Comparison with Numerical Results}
\label{sec:Results}
To assess the reliability of our analytical model, we also numerically calculate the CSM profile using the radiation hydrodynamics code developed by \citet{Kuriyama20a}. We fit the numerical profile with the function in equation (\ref{eq:rho_CSM}) using the least-squares method, with $(r_*,\rho_*,y)$ as fitting parameters.

We prepare three hydrogen-rich progenitors, two RSGs with zero-age main sequence (ZAMS) mass of $15$ and $20M_\odot$, and a BSG with ZAMS mass of $20M_\odot$. The stars are generated by the MESA code \citep{Paxton11,Paxton13,Paxton15,Paxton18,Paxton19} version 12778, using the \verb|example_make_pre_ccsn| test suite. For the former two RSGs the default parameters are used and solar metallicity was assumed. To obtain the BSG progenitor we change the metallicity to $10^{-2}Z_\odot$, and adopt a thermohaline efficiency parameter of $1$ for semiconvection with the Ledoux criterion (the default value is $0$; for discussion of this parameter see \cite{Paxton13} and references therein). As the structure of the envelope does not change in the last decades of its life, we use the stars evolved up to core-collapse for the input of the simulation. The properties of the progenitors are summarized in Table \ref{table:progenitors}.

As in \citet{Kuriyama20a}, we inject some energy at the base of the hydrogen envelope and follow the response of the envelope. The injected energy is comparable to the envelope's binding energy $E_{\rm bind}$ and thus parameterized as $E_{\rm inj}=f_{\rm inj}E_{\rm bind}$. We adopt the values $f_{\rm inj}=\{0.3, 0.5\}$, which results in a range of CSM masses in line with what is usually considered to reproduce Type IIn SNe \citep{Kuriyama20a}.

\begin{figure*}
 \centering
 \includegraphics[width=\linewidth]{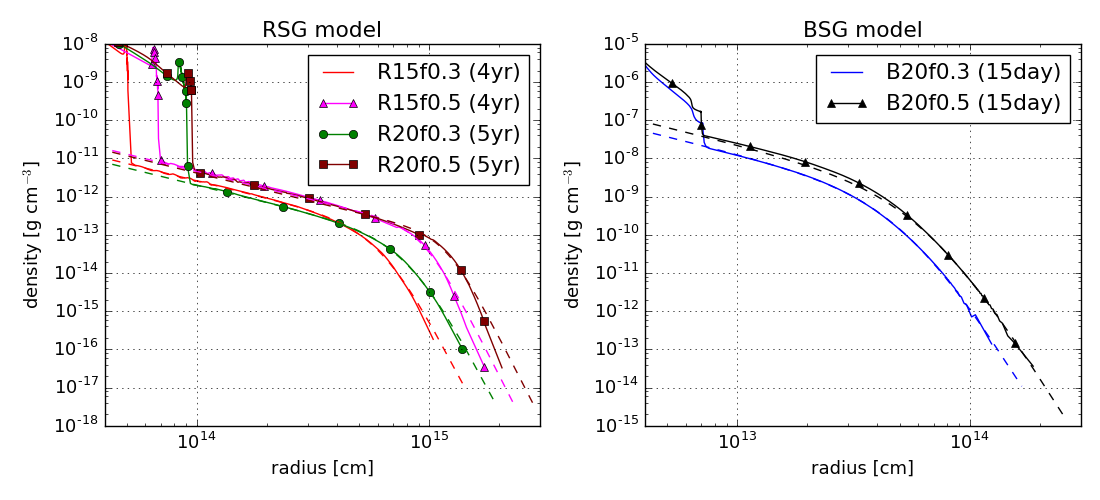}
\caption{CSM density profiles obtained by numerical simulations (solid lines) compared with our analytical fit (dashed lines). The left panel is for the RSG progenitors, while the right panel is for the BSG progenitors.}
 \label{fig:density_profile}
 \end{figure*}

The simulation gives a density profile of the entire matter that was originally the hydrogen-rich envelope of the progenitor. This means that we need to define the interface between the star (which eventually becomes the SN ejecta) and the CSM. As can be seen in the density profiles of Figure \ref{fig:density_profile}, there is ubiquitously a density jump, which can be ascribed to the infalling CSM crashing the stellar envelope. We thus assume that the smooth profile outside this density jump is the CSM, and fit this region with the analytical profile. The number of cells used for the fit is of order 1000. As $r$ and $\rho$ span more than an order of magnitude, we fit the function $\log\rho(\log(r))$ to ensure contribution among the entire fitting region.

The time-dependence of the profile can be easily derived once the fluid element at $r=r_*$ starts to expand homologously. During this phase the value of the curvature parameter $y$ is expected to be unchanged. This is because $y$ is governed only by the CSM in the vicinity of $r=r_*$ that should also be homologously expanding. Taking into account  the time evolution of the CSM mass, the parameters evolve as $r_*\propto t$, and $\rho_*$ as given in equation (\ref{eq:rhostar_evolution}) in Appendix \ref{sec:homologous}. This homologous phase sets in from when the specific kinetic energy of the CSM at the transition radius, $\approx r_*^2/2t^2$, is much larger than the gravitational potential at that radius, $GM_*/r_*$. Setting a parameter $v_*=r_*/t$, which is constant of time in the homologous phase, the ratio of the two energies is
\begin{eqnarray}
\frac{v_*^2/2}{GM_*/(tv_*)} \sim 12\left(\frac{t}{1\ {\rm yr}}\right)\left(\frac{M_*}{10M_\odot}\right)^{-1}\left(\frac{v_*}{10^7\ {\rm cm\ s^{-1}}}\right)^3.
\end{eqnarray}
After this ratio becomes much larger than 1, the above simple scaling can be used when considering the future evolution of the profile. Thus we stop the simulations at the time $t_{\rm stop}$ (defined as the epoch from energy injection) when this ratio is $\gtrsim 5$, for the sake of computational cost. The value of $t_{\rm stop}$ for BSGs is much shorter than that for RSGs due to the larger $v_*$ (see Table \ref{table:results}).

Table \ref{table:results} shows our fitting results, and Figure \ref{fig:density_profile} shows the corresponding fits. We find that our analytical profile produces good fits to the numerical results, although we find discrepancies in the tails of the RSG models. A similar discrepancy is seen in \citet{Matzner99}, which they ascribe to the presence of the superadiabatic gradient close to the surface of RSGs. We presume the same physics operate in our models, but this difference in the tail, which carries negligible mass in the CSM, will likely not affect the observational properties of interaction-powered SNe.

\begin{table*}
\centering
\begin{tabular}{ccc|cc|cccc}
Model & $E_{\rm inj}$ [$10^{47}$erg] & $t_{\rm stop}$ & $R_{\rm in, CSM}$ [cm] & $M_{\rm CSM}$ [$M_\odot$] & $r_*$ [cm] & $\rho_*$ [g cm$^{-3}$] & $y$ \\ \hline
R15f0.3 & 1.4 & 4 years & $5.0\times 10^{13}$ & 0.11 & $6.4\times 10^{14}$ & $2.7\times 10^{-14}$ & 2.6\\
R15f0.5 & 2.4 & 4 years & $7.0\times 10^{13}$ & 0.49 & $9.5\times 10^{14}$ & $4.9\times 10^{-14}$ & 1.7\\
R20f0.3 & 2.2 & 5 years & $9.1\times 10^{13}$ & 0.13 & $8.5\times 10^{14}$ & $1.1\times 10^{-14}$ & 2.8 \\
R20f0.5 & 3.7 & 5 years & $9.5\times 10^{13}$ & 0.62 & $1.2\times 10^{15}$ & $3.2\times 10^{-14}$ & 1.7 \\
B20f0.3 & 54 & 15 days & $7.2\times 10^{12}$ & $0.30$ & $5.5\times 10^{13}$ & $8.7\times 10^{-11}$ & $3.6$\\
B20f0.5 & 91 & 15 days & $7.0\times 10^{12}$ & $0.79$& $6.7\times 10^{13}$ & $9.5\times 10^{-11}$ & $3.8$\\
\end{tabular}
\caption{Models of the CSM, and results of fitting for the numerical CSM profile with our analytical model. The first five columns are model name, injected energy, time that we stop our simulation, and the innermost radius and total mass of the region that we define as CSM. The last three columns are results of the least-squares fitting, with fitted parameters ($r_*,\rho_*,y$).}
\label{table:results}
\end{table*}

\section{Discussions}
\label{sec:Discussion}
\subsection{Implication for the Bolometric Light Curve}
We consider the implications of our density profile model, by comparing our profile with the commonly adopted ``wind profile" of $\rho\propto r^{-2}$. As an example, we compare the bolometric light curves from ejecta-CSM interaction, using a numerical model by \citet{Takei20}.

\begin{figure*}
 \centering
 \includegraphics[width=\linewidth]{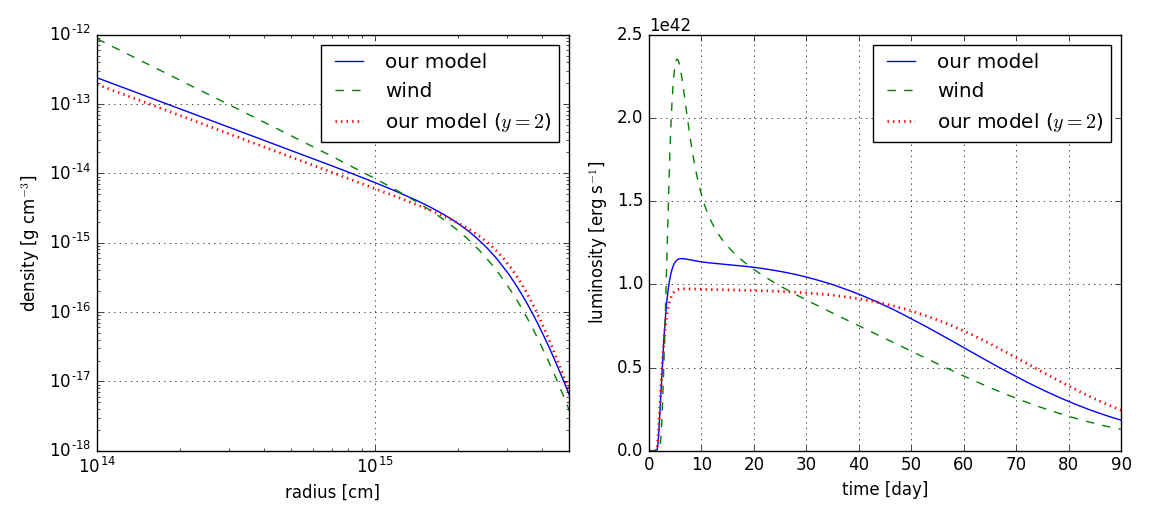}
\caption{Comparison of our profile with a wind profile at the inner edge. The left panel shows the assumed CSM profiles for input to the light curve model, all having a total mass of $0.13M_\odot$ in the range $10^{14}$ cm $<r<10^{16}$ cm. The right panel shows the bolometric light curves for the profiles.}
 \label{fig:lightcurve}
 \end{figure*}

We consider a CSM of our analytical profile of equation (\ref{eq:rho_CSM}) with parameters $r_*=3.4\times 10^{15}$ cm, $y=2.8$ and total CSM mass of $0.13M_\odot$, and compare this against the wind profile. Our parameters correspond to the R20f0.3 model, but assuming that the energy was injected 20 years before core collapse. For the wind profile we use a similar function as equation (\ref{eq:rho_CSM}), but change the inner power-law index to asymptotically be $-2$ instead of $-1.5$. We use the same $r_*$ and $y$ for the two profiles, but use a different value for $\rho_*$ so that the two CSMs have the same total mass. We show the density profiles in the left panel of Figure \ref{fig:lightcurve}.

For the ejecta we adopt mass and energy of $M_{\rm ej}=15$ $M_\odot$ and $E_{\rm ej}=10^{51}$ erg respectively. We assume homologous ejecta with a density profile \citep{Matzner99} 
\begin{eqnarray}
\rho_{\rm ej} (r,t)
&=& \left\{ \begin{array}{ll}
t^{-3}\left[r/(gt)\right]^{-n} & (r/t > v_t),\\
t^{-3}(v_t/g)^{-n} \left[r/(tv_t)\right]^{-\delta}  & (r/t < v_t),
\end{array}\right.
\label{eq:rho_ej}
\end{eqnarray}
where we adopt $n=12$ and $\delta=1$. Factors $g, v_t$ are constants determined from $M_{\rm ej}$ and $E_{\rm ej}$. We refer to \citet{Takei20} for the details of the light curve model.

The right panel of Figure \ref{fig:lightcurve} shows the resulting bolometric light curves. We find that the bolometric light curve from our shallower profile displays a much flatter shape than that from the wind profile. The flat light curve is due to the constant kinetic energy dissipation rate for $n=12$ and CSM with power-law index $-1.5$ \citep{Moriya_et_al_13}. However the plateau may not be universal, since the light curve would depend on the diffusion timescale in the CSM and the time for the shock to reach $r_*$. It would be interesting to do a more exhaustive parameter study and compare the light curve models with observations of interaction-powered SNe. We plan to do this in future work.

\subsection{A Guide for Modelling our CSM Profile}
An important application of our model would be to compare with observations of interaction-powered SNe to extract the parameters of the CSM. As observed data also depend on other parameters such as the ejecta's kinetic energy and radiation conversion efficiency, we are usually only interested in a rough inference of the radius and mass of the CSM, which relate to the epoch and strength of the mass loss. Here we show that in this case only $r_*$ and $\rho_*$ are important, and $y$ is not much an important factor.

In Figure \ref{fig:lightcurve} we also show the difference in the light curve if we change from our fiducial case of $y=2.8$ (solid line) to $y=2$ (dotted line), with the total CSM mass again being the same. We find that while the shape of the light curve becomes slightly modified, the overall luminosity and timescale is similar. We conclude that while varying $y$ can make slight changes to observables, it is relatively less important than the other two parameters $r_*, \rho_*$.

The total CSM mass calculated from the analytical profile is
$M_{\rm CSM} \approx \int_0^\infty 4\pi r^2\rho_{\rm CSM} dr
\equiv 4\pi r_*^3\rho_* \eta(y)$,
where
\begin{eqnarray}
\eta(y)&\equiv& \int_0^\infty dx\, x^2\left[\frac{x^{1.5/y}+x^{n_{\rm max}}/y}{2}\right]^{(-y)},\ x\equiv r/r_*.
\label{eq:MCSM_tot}
\end{eqnarray}
We find that in the range $1\leq y\leq 4$, $\eta$ varies by a factor of $\sim 3$. Fixing $y\approx2$ ($y\approx 4$) for RSGs (BSGs), and modelling the CSM by just two parameters ($r_*,\rho_*$ or $r_*,M_{\rm CSM}$) should be sufficient if one is interested in order-of-magnitude estimates of the extent and mass of the CSM. We therefore recommend the density profiles
\begin{eqnarray}
\rho_{\rm CSM}(r) 
&=& \left\{ \begin{array}{ll}
\rho_* \left[\frac{(r/r_*)^{0.75}+(r/r_*)^{6}}{2}\right]^{-2} & ({\rm RSGs}) ,\\
\rho_* \left[\frac{(r/r_*)^{0.375}+(r/r_*)^{2.5}}{2}\right]^{-4} & ({\rm BSGs}).
\end{array}\right.
\label{eq:rho_CSM_simplified}
\end{eqnarray}
This two-parameter modelling is complementary to the commonly used power-law profile $\rho_{\rm CSM}(r)=qr^{-s}$ that also has two parameters: index $s$ and normalization $q$.


\section{Conclusions}
\label{sec:Conclusion}

In this work we have derived an analyical density profile of a CSM created from a single mass eruption event years to decades before core-collapse. We find that the density profile is well described by a double power-law, reflecting the bound and unbound limits of the CSM under the influence of gravitational pull from the central star. Using numerical simulations of \citet{Kuriyama20a}, we verified that our profile is in good agreement with that numerically obtained from radiation hydrodynamical calculations.

We have shown that the radius and density at the transition are the two important parameters of our model that one should vary when trying to reproduce observations. We encourage future works attempting to model observations (e.g. light curves, spectra) of interaction-powered SNe to include our profile as input if possible.

The main conclusion of our model is that the inner part of the double power-law density profile follows $\rho\propto r^{-1.5}$, shallower than the wind profile. This flat profile is consistent with that reported for a Type IIn SN 2006jd \citep{Chandra12}. An independent study that modelled the optical light curve of SN 2006jd \citep{Moriya14} confirms this, while it finds a steeper CSM for many of the other samples. We note however that light curve modelling can be subject to various systematic uncertainties, with different models disagreeing on the inferred density profile (see e.g. discussion in \cite{Takei20} for SN 2005kj).

We mention a few possible caveats of our model. First our model assumes single mass eruption, while  multiple mass eruptions are observed for some Type IIn SNe, the representative being SN 2009ip \citep{Smith10,Pastorello13}. Though Type IIn SNe with signatures of multiple mass eruptions with a short interval (within years) are likely rare \citep{Nyholm20}, for such cases the density profile can be different from our simple profile (see e.g. Figure 12 of \cite{Kuriyama21}).

Second, our model well reproduces CSM that expands homologously at the outermost radius. For lower $f_{\rm inj}$, or shorter interval from expansion to core-collapse (comparable to or less than the progenitor's dynamical timescale), the outermost CSM would not be homologous and the profile can be more complex. However, these cases would lead to much smaller extent of the CSM and/or much lighter CSM mass, which may not reproduce the features of the CSM in observed interaction-powered SNe. 

\begin{ack}
We thank the anonymous referee for many insightful comments that greatly improved the manuscript. DT is supported by the Advanced Leading Graduate Course for Photon Science (ALPS) at the University of Tokyo. 
YT is supported by the RIKEN Junior Research
Associate Program.
This work is also supported by JSPS KAKENHI Grant Numbers JP19J21578, JP16H06341, JP20H05639, MEXT, Japan.
\end{ack}

\begin{appendix}

\section{Homologous Expansion Phase}
\label{sec:homologous}
After the CSM at $r=r_*$ reaches the homologous phase, the velocity $v_*$ and curvature parameter $y$ are expected to be constant. At the innermost region, the CSM plunges into the star with a negative velocity, which reduces the mass of the CSM. In this section we first derive the total CSM mass and the density normalization parameter $\rho_*$ as a function of time, and compare this time evolution with the numerical results.

The mass inflow (loss) rate of CSM from the innermost radius $R_{\rm in, CSM}$ is
\begin{eqnarray}
\frac{dM_{\rm CSM}}{dt}&=&(4\pi r^2\rho v)_{r=R_{\rm in, CSM}} \nonumber \\
&\approx& -4\pi\sqrt{2GM_*R_{\rm in, CSM}^3}\times \rho(R_{\rm in, CSM}),
\end{eqnarray}
where we have used the free-fall approximation $v\approx -\sqrt{2GM_*/r}$. From the analytical profile
\begin{eqnarray}
\rho(R_{\rm in, CSM}) &=& \rho_*\left[\frac{(R_{\rm in, CSM}/r_*)^{1.5/y}+(R_{\rm in, CSM}/r_*)^{n_{\rm max}/y}}{2}\right]^{-y} \nonumber \\
&\approx& 2^y\rho_*\left(\frac{R_{\rm in, CSM}}{r_*}\right)^{-1.5},
\end{eqnarray}
where we used $(R_{\rm in, CSM}/r_*)^{n_{\rm max}-1.5}\ll 1$. Using the relation $M_{\rm CSM}\approx 4\pi r_*^3\rho_*\eta(y)$ in Section \ref{sec:Discussion} with $\eta(y)$ defined in equation (\ref{eq:MCSM_tot}), we obtain the following differential equation
\begin{eqnarray}
\frac{dM_{\rm CSM}}{dt}&\approx& -\frac{2^{y+0.5}}{\eta(y)}\sqrt{\frac{GM_*}{v_*^3}}M_{\rm CSM}t^{-3/2}.
\end{eqnarray}
The solution to this equation is
\begin{eqnarray}
M_{\rm CSM} = M_{\rm stop}\exp\left[\frac{2^{y+1.5}}{\eta(y)}\sqrt{\frac{GM_*}{v_*^3}}(t^{-1/2}-t_{\rm stop}^{-1/2})\right],
\end{eqnarray}
where $t_{\rm stop}$ is the time when the fluid at $r=r_*$ enters a homologous expansion phase (see also section 3 and Table 2), the constant $M_{\rm stop}$ is the total CSM mass at $t=t_{\rm stop}$. We note that while this solution diverges at $t\to 0$, it is not valid at $t\to 0$ because the homologous phase is yet to be achieved. This expression can be rewritten as $M_{\rm CSM} \propto \exp\left[(t/t_c)^{-1/2}-(t_{\rm stop}/t_c)^{-1/2}\right]$, where
\begin{eqnarray}
t_c = \frac{2^{2y+3}}{\eta(y)^2}\frac{GM_*}{v_*^3} \sim 1\ {\rm  yr} \left(\frac{M_*}{10M_\odot}\right)\left(\frac{v_*}{100\ {\rm km\ s^{-1}}}\right)^{-3}.
\end{eqnarray}
The latter scaling is for a reference value $y=2$. The numerical factor $2^{2y+3}/\eta(y)^2$ is a monotonically increasing function of $y$, ranging from $23$ to $88$ for $1\leq y\leq 4$ and $n_{\rm max}=12$. The time-dependence of $\rho_*$ is then
\begin{eqnarray}
\rho_*(t) \propto t^{-3}\exp\left[\left(\frac{t}{t_c}\right)^{-1/2}\right].
\label{eq:rhostar_evolution}
\end{eqnarray}

To test this solution, we carry out the numerical simulation up to $10$ years for Model R15f0.3. Then in addition to $t=4$ years presented in the main text, we fit the CSM profile at $t=2, 7, 10$ years, with the same fitting parameters $(r_*,\,\rho_*,\,y)$.

\begin{figure*}[ht]
 \centering
 \includegraphics[width=\linewidth]{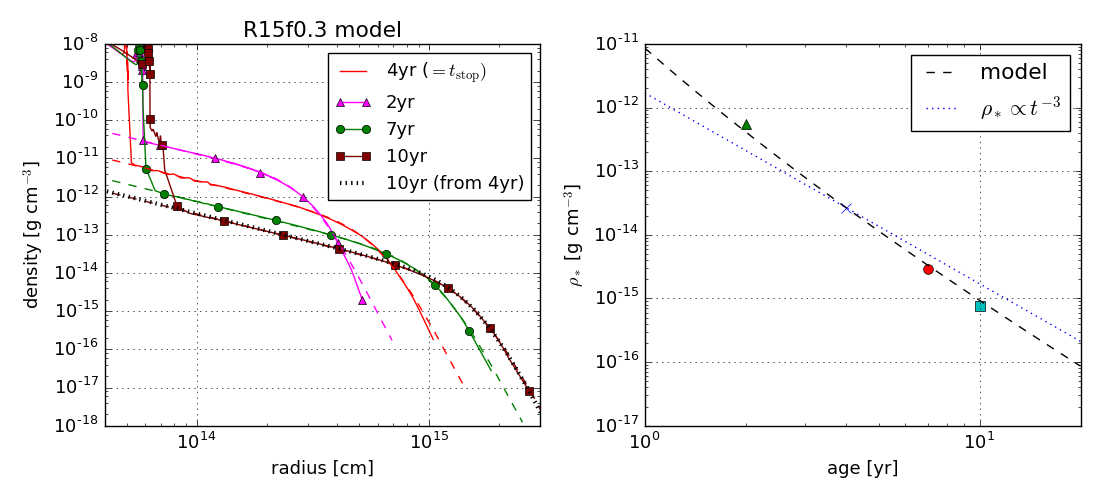}
\caption{Time evolution of the CSM density profile (Left panel). The density profiles obtained by numerical simulations (solid lines) compared with our analytical fit (dashed lines) for various ages from 2 to 10 years. The black dotted line is the CSM profile at $t=10$ years obtained from using the  parameters fitted with equation (\ref{eq:rho_CSM_simplified}) for $t=4$ years and the scaling of $r_*$ and $\rho_*$ formulated in Appendix \ref{sec:homologous}. (Right panel) Values of $\rho_*$ obtained from fitting for the four values of $t$, compared with the analytical formula in equation (\ref{eq:rhostar_evolution}) shown as a dashed line. The normalization of the two lines is set to cross the point at $t=4$ years.}
 \label{fig:csmfit_evolution}
 \end{figure*}

The fitting results are shown in Figure \ref{fig:csmfit_evolution} and Table \ref{table:results_appendix}. The analytical profile fits well the numerically obtained profile at all times. We find that the value of $y$ is nearly converged at $t=t_{\rm stop}=4$ yr, with future change of within only $\approx6$\%. We note that this small difference in $y$ has a tiny effect on the observables, such as the light curve discussed in Section \ref{sec:Discussion}.

Using the fitting parameters at $t=4$ years and equation (\ref{eq:rhostar_evolution}), we analytically calculate the density profile at $t=10$ years. The value of $t_c$ for this model is obtained from the fitting parameters as $t_c\approx 11$ years. The result of using this scaling is shown as black dotted lines in the left panel of Figure \ref{fig:csmfit_evolution}. We find that the density profile well matches the numerical results, which validates the analytical evolution and our assumption on transition to homologous flow at $t=t_{\rm stop}$. The right panel of Figure \ref{fig:csmfit_evolution} shows the value of $\rho_*$ at the four epochs, compared with the analytical curve of equation (\ref{eq:rhostar_evolution}). We confirm that the analytical model gives a good match to the numerical results.

\begin{table}
\centering
\begin{tabular}{cc|ccc}
$t$ [yr] & $R_{\rm in, CSM}$ [cm] & $r_*$ [cm] & $\rho_*$ [g cm$^{-3}$] & $y$ \\ \hline
2 & $5.0\times 10^{13}$ & $3.1\times 10^{14}$ & $5.5\times 10^{-13}$ & 2.08\\
4 & $5.9\times 10^{13}$ & $6.4\times 10^{14}$ & $2.7\times 10^{-14}$ & 2.61\\
7 & $6.6\times 10^{13}$ & $1.1\times 10^{15}$ & $2.9\times 10^{-15}$ & 2.77\\
10 & $8.2\times 10^{13}$ & $1.6\times 10^{15}$ & $7.7\times 10^{-16}$ & 2.78\\
\end{tabular}
\caption{Results of fitting for the numerical CSM profile with our analytical model for different epochs for the R15f0.3 model. The columns are age, the innermost radius that we define as CSM, and the fitted parameters ($r_*,\rho_*,y$). The value for $t=4$ yr is the same as that in Table \ref{table:results}.}
\label{table:results_appendix}
\end{table}

\end{appendix}

\bibliographystyle{apj} 
\bibliography{CSM}

\begin{thebibliography}{31}
\expandafter\ifx\csname natexlab\endcsname\relax\def\natexlab#1{#1}\fi

\bibitem[{{Bondi}(1952)}]{Bondi52}
{Bondi}, H. 1952, \mnras, 112, 195

\bibitem[{{Bruch} {et~al.}(2020){Bruch}, {Gal-Yam}, {Schulze}, {Yaron}, {Yang},
  {Soumagnac}, {Rigault}, {Strotjohann}, {Ofek}, {Sollerman}, {Masci},
  {Barbarino}, {Ho}, {Fremling}, {Perley}, {Nordin}, {Cenko}, {Adams},
  {Adreoni}, {Bellm}, {Blagorodnova}, {Bulla}, {Burdge}, {De}, {Dhawan},
  {Drake}, {Duev}, {Dugas}, {Graham}, {Graham}, {Jencson}, {Karamehmetoglu},
  {Kasliwal}, {Kim}, {Kulkarni}, {Kupfer}, {Mahabal}, {Miller}, {Prince},
  {Riddle}, {Sharma}, {Smith}, {Taddia}, {Taggart}, {Walters}, \&
  {Yan}}]{Bruch20}
{Bruch}, R.~J., et~al. 2020, arXiv e-prints, arXiv:2008.09986

\bibitem[{{Chandra} {et~al.}(2012){Chandra}, {Chevalier}, {Chugai}, {Fransson},
  {Irwin}, {Soderberg}, {Chakraborti}, \& {Immler}}]{Chandra12}
{Chandra}, P., {Chevalier}, R.~A., {Chugai}, N., {Fransson}, C., {Irwin},
  C.~M., {Soderberg}, A.~M., {Chakraborti}, S., \& {Immler}, S. 2012, \apj,
  755, 110

\bibitem[{{Chatzopoulos} {et~al.}(2012){Chatzopoulos}, {Wheeler}, \&
  {Vinko}}]{Chatzopoulos12}
{Chatzopoulos}, E., {Wheeler}, J.~C., \& {Vinko}, J. 2012, \apj, 746, 121

\bibitem[{{Chevalier} \& {Irwin}(2011)}]{Chevalier11}
{Chevalier}, R.~A., \& {Irwin}, C.~M. 2011, \apjl, 729, L6

\bibitem[{{Ginzburg} \& {Balberg}(2012)}]{Ginzburg12}
{Ginzburg}, S., \& {Balberg}, S. 2012, \apj, 757, 178

\bibitem[{{Grasberg} \& {Nadyozhin}(1987)}]{Grasberg87}
{Grasberg}, E.~K., \& {Nadyozhin}, D.~K. 1987, \azh, 64, 1199

\bibitem[{{Kuriyama} \& {Shigeyama}(2020)}]{Kuriyama20a}
{Kuriyama}, N., \& {Shigeyama}, T. 2020, \aap, 635, A127

\bibitem[{{Kuriyama} \& {Shigeyama}(2021)}]{Kuriyama21}
---. 2021, \aap, 646, A118

\bibitem[{{Matzner} \& {McKee}(1999)}]{Matzner99}
{Matzner}, C.~D., \& {McKee}, C.~F. 1999, \apj, 510, 379

\bibitem[{{Moriya} {et~al.}(2011){Moriya}, {Tominaga}, {Blinnikov}, {Baklanov},
  \& {Sorokina}}]{Moriya11}
{Moriya}, T., {Tominaga}, N., {Blinnikov}, S.~I., {Baklanov}, P.~V., \&
  {Sorokina}, E.~I. 2011, \mnras, 415, 199

\bibitem[{{Moriya} {et~al.}(2013){Moriya}, {Maeda}, {Taddia}, {Sollerman},
  {Blinnikov}, \& {Sorokina}}]{Moriya_et_al_13}
{Moriya}, T.~J., {Maeda}, K., {Taddia}, F., {Sollerman}, J., {Blinnikov},
  S.~I., \& {Sorokina}, E.~I. 2013, \mnras, 435, 1520

\bibitem[{{Moriya} {et~al.}(2014){Moriya}, {Maeda}, {Taddia}, {Sollerman},
  {Blinnikov}, \& {Sorokina}}]{Moriya14}
---. 2014, \mnras, 439, 2917

\bibitem[{{Moriya} {et~al.}(2018){Moriya}, {Sorokina}, \&
  {Chevalier}}]{Moriya18}
{Moriya}, T.~J., {Sorokina}, E.~I., \& {Chevalier}, R.~A. 2018, \ssr, 214, 59

\bibitem[{{Morozova} {et~al.}(2018){Morozova}, {Piro}, \&
  {Valenti}}]{Morozova18}
{Morozova}, V., {Piro}, A.~L., \& {Valenti}, S. 2018, \apj, 858, 15

\bibitem[{{Murase} {et~al.}(2019){Murase}, {Franckowiak}, {Maeda}, {Margutti},
  \& {Beacom}}]{Murase19}
{Murase}, K., {Franckowiak}, A., {Maeda}, K., {Margutti}, R., \& {Beacom},
  J.~F. 2019, \apj, 874, 80

\bibitem[{{Nyholm} {et~al.}(2020){Nyholm}, {Sollerman}, {Tartaglia}, {Taddia},
  {Fremling}, {Blagorodnova}, {Filippenko}, {Gal-Yam}, {Howell},
  {Karamehmetoglu}, {Kulkarni}, {Laher}, {Leloudas}, {Masci}, {Kasliwal},
  {Mor{\r{a}}}, {Moriya}, {Ofek}, {Papadogiannakis}, {Quimby}, {Rebbapragada},
  \& {Schulze}}]{Nyholm20}
{Nyholm}, A., et~al. 2020, \aap, 637, A73

\bibitem[{{Pastorello} {et~al.}(2013){Pastorello}, {Cappellaro}, {Inserra},
  {Smartt}, {Pignata}, {Benetti}, {Valenti}, {Fraser}, {Tak{\'a}ts}, {Benitez},
  {Botticella}, {Brimacombe}, {Bufano}, {Cellier-Holzem}, {Costado}, {Cupani},
  {Curtis}, {Elias-Rosa}, {Ergon}, {Fynbo}, {Hambsch}, {Hamuy}, {Harutyunyan},
  {Ivarson}, {Kankare}, {Martin}, {Kotak}, {LaCluyze}, {Maguire}, {Mattila},
  {Maza}, {McCrum}, {Miluzio}, {Norgaard-Nielsen}, {Nysewander}, {Ochner},
  {Pan}, {Pumo}, {Reichart}, {Tan}, {Taubenberger}, {Tomasella}, {Turatto}, \&
  {Wright}}]{Pastorello13}
{Pastorello}, A., et~al. 2013, \apj, 767, 1

\bibitem[{{Paxton} {et~al.}(2011){Paxton}, {Bildsten}, {Dotter}, {Herwig},
  {Lesaffre}, \& {Timmes}}]{Paxton11}
{Paxton}, B., {Bildsten}, L., {Dotter}, A., {Herwig}, F., {Lesaffre}, P., \&
  {Timmes}, F. 2011, \apjs, 192, 3

\bibitem[{{Paxton} {et~al.}(2013){Paxton}, {Cantiello}, {Arras}, {Bildsten},
  {Brown}, {Dotter}, {Mankovich}, {Montgomery}, {Stello}, {Timmes}, \&
  {Townsend}}]{Paxton13}
{Paxton}, B., et~al. 2013, \apjs, 208, 4

\bibitem[{{Paxton} {et~al.}(2015){Paxton}, {Marchant}, {Schwab}, {Bauer},
  {Bildsten}, {Cantiello}, {Dessart}, {Farmer}, {Hu}, {Langer}, {Townsend},
  {Townsley}, \& {Timmes}}]{Paxton15}
---. 2015, \apjs, 220, 15

\bibitem[{{Paxton} {et~al.}(2018){Paxton}, {Schwab}, {Bauer}, {Bildsten},
  {Blinnikov}, {Duffell}, {Farmer}, {Goldberg}, {Marchant}, {Sorokina},
  {Thoul}, {Townsend}, \& {Timmes}}]{Paxton18}
---. 2018, \apjs, 234, 34

\bibitem[{{Paxton} {et~al.}(2019){Paxton}, {Smolec}, {Schwab}, {Gautschy},
  {Bildsten}, {Cantiello}, {Dotter}, {Farmer}, {Goldberg}, {Jermyn}, {Kanbur},
  {Marchant}, {Thoul}, {Townsend}, {Wolf}, {Zhang}, \& {Timmes}}]{Paxton19}
---. 2019, \apjs, 243, 10

\bibitem[{{Smith} {et~al.}(2010){Smith}, {Miller}, {Li}, {Filippenko},
  {Silverman}, {Howard}, {Nugent}, {Marcy}, {Bloom}, {Ghez}, {Lu}, {Yelda},
  {Bernstein}, \& {Colucci}}]{Smith10}
{Smith}, N., et~al. 2010, \aj, 139, 1451

\bibitem[{{Soker}(2021)}]{Soker21}
{Soker}, N. 2021, \apj, 906, 1

\bibitem[{{Suzuki} {et~al.}(2020){Suzuki}, {Moriya}, \& {Takiwaki}}]{Suzuki20}
{Suzuki}, A., {Moriya}, T.~J., \& {Takiwaki}, T. 2020, \apj, 899, 56

\bibitem[{{Svirski} {et~al.}(2012){Svirski}, {Nakar}, \& {Sari}}]{Svirski12}
{Svirski}, G., {Nakar}, E., \& {Sari}, R. 2012, \apj, 759, 108

\bibitem[{{Takei} \& {Shigeyama}(2020)}]{Takei20}
{Takei}, Y., \& {Shigeyama}, T. 2020, \pasj, 72, 67

\bibitem[{{Tsuna} {et~al.}(2020){Tsuna}, {Ishii}, {Kuriyama}, {Kashiyama}, \&
  {Shigeyama}}]{Tsuna20}
{Tsuna}, D., {Ishii}, A., {Kuriyama}, N., {Kashiyama}, K., \& {Shigeyama}, T.
  2020, \apjl, 897, L44

\bibitem[{{Tsuna} {et~al.}(2019){Tsuna}, {Kashiyama}, \& {Shigeyama}}]{Tsuna19}
{Tsuna}, D., {Kashiyama}, K., \& {Shigeyama}, T. 2019, \apj, 884, 87

\bibitem[{{Yaron} {et~al.}(2017){Yaron}, {Perley}, {Gal-Yam}, {Groh}, {Horesh},
  {Ofek}, {Kulkarni}, {Sollerman}, {Fransson}, {Rubin}, {Szabo}, {Sapir},
  {Taddia}, {Cenko}, {Valenti}, {Arcavi}, {Howell}, {Kasliwal}, {Vreeswijk},
  {Khazov}, {Fox}, {Cao}, {Gnat}, {Kelly}, {Nugent}, {Filippenko}, {Laher},
  {Wozniak}, {Lee}, {Rebbapragada}, {Maguire}, {Sullivan}, \&
  {Soumagnac}}]{Yaron17}
{Yaron}, O., et~al. 2017, Nature Physics, 13, 510

\end{thebibliography}

\end{document}